\documentclass[12pt]{article}
\newcommand{\Ir}{Z\!\!\!Z}
\newcommand{\idty}{{\leavevmode{\rm 1\mkern -5.4mu I}}}
\newcommand{\Ibb}[1]{ {\rm I\ifmmode\mkern
            -3.6mu\else\kern -.2em\fi#1}}
\newcommand{\ibb}[1]{\leavevmode\hbox{\kern.3em\vrule
     height 1.2ex depth -.3ex width .2pt\kern-.3em\rm#1}}
\newcommand{\Cx}{{\ibb C}}
\newcommand{\Rl}{{\Ibb R}}
\newcommand{\Nl}{{\Ibb N}}

\newcommand{\be}{\begin{equation}}
\newcommand{\ee}{\end{equation}}
\newcommand{\bea}{\begin{eqnarray}}
\newcommand{\eea}{\end{eqnarray}}
\newcommand{\A}{{\cal A}}
\renewcommand{\O}{\mit\Omega}
\newcommand{\pa}{\partial}
\renewcommand{\d}{{\rm d}}
\newcommand{\oa}{\otimes_\A}
\newcommand{\oL}{\otimes_L}

\begin{document}

\title{PSEUDO-RIEMANNIAN METRICS IN MODELS BASED ON NONCOMMUTATIVE GEOMETRY}
\date{  }
\author{A. Dimakis \\ Department of Mathematics, University of the Aegean \\
        GR-83200 Karlovasi, Samos, Greece \\ dimakis@aegean.gr
        \\[2ex]
        F. M\"uller-Hoissen \\ Max-Planck-Institut f\"ur Str\"omungsforschung \\
        Bunsenstrasse 10, D-37073 G\"ottingen, Germany \\
        fmuelle@gwdg.de }

\maketitle

\begin{abstract}
Several examples and models based on noncommutative differential calculi on commutative algebras indicate that a metric should be regarded as an element of the left-linear tensor product of the space of 1-forms with itself. We show how the metric compatibility condition with a linear connection generalizes to this framework.
\end{abstract}

\section{Introduction}
In ordinary differential geometry there are several equivalent ways to introduce the concept of a metric. It should not be surprising, however, that corresponding generalizations to the huge framework of noncommutative geometry in general lead to somehow inequivalent structures. Some definition of a metric in noncommutative geometry is easily chosen and calculations performed with it. What is often missing, however, is a serious application which demonstrates its usefulness outside the single point which corresponds to ordinary differential geometry in the set of noncommutative differential geometries. We do not believe that a particular definition of a metric will finally be singled out by general arguments. Rather, we expect that a convenient definition of a metric will depend on the area of applications which one has in mind, and the relations between different definitions may turn out to be extremely complicated. 

In section 2 we briefly recall some metric definitions in noncommutative geometry and add to it a new one which, however, is restricted to the case of (noncommutative) differential calculi on {\em commutative} algebras, which includes the case of discrete spaces. Section 3 provides some examples and models in which this metric definition shows up. In section 4 we introduce compatibility of such a metric with a linear connection. Section 5 contains some conclusions.

\section{General setting for noncommutative pseudo-Riemannian geometry}

Let $\A$ be an associative algebra (over $\Cx$ or $\Rl$) with unit element $\idty$.  
A {\em graded algebra} over $\A$ is a $\Ir$-graded associative algebra 
$ \O (\A) = \bigoplus_{r \geq 0} \O^r (\A)$ 
where $\O^0(\A) = \A$. 
A {\em differential calculus} over $\A$ consists of a graded algebra $\O(\A)$ over $\A$ and a linear\footnote{If $\A$ is an algebra over $\Cx$ ($\Rl$), a {\em linear} map is linear over $\Cx$ ($\Rl$).} map $ \d \, : \,  \O^r (\A) \rightarrow \O^{r+1}(\A)$ with the properties
\bea
       \d^2 = 0 \, , \qquad
       \d (w \, w') = (\d w) \, w' + (-1)^r \, w \, \d w'                 
\eea
where $w \in \O^r(\A)$ and $w' \in \O (\A)$. We also require $\idty \, w = w \, \idty = w$ for all elements $w \in \O (\A)$. The identity $\idty \idty = \idty$ then implies $ \d \idty = 0 $. In the following, we simply write $\O$ instead of $\O(\A)$.
\vskip.2cm

A (left $\A$-module) {\em linear connection} is a linear map
$\nabla \, : \, \O^1 \rightarrow \O^1 \oa \O^1$ such that
\be
    \nabla (f \, \alpha) = \d f \oa \alpha + f \, \nabla \alpha
\ee
for $f \in \A$ and $\alpha \in \O^1$. It extends to a linear map
$\nabla \, : \; \O \oa \O^1 \rightarrow \O \oa \O^1$ via
\be
   \nabla (w \oa \alpha) = \d w \oa \alpha + (-1)^r \, w \, \nabla \alpha
   \qquad  w \in \O^r, \; \alpha \in \O^1  \; .
\ee
The {\em curvature} of the linear connection $\nabla$ is the map
${\cal R} = - \nabla^2$ and the {\em torsion} 
$\Theta \, : \, \O \oa \O^1 \rightarrow \O$ is defined as
$ \Theta = \d \circ \pi - \pi \circ \nabla $
where $\pi$ is the projection $\O \oa \O^1 \rightarrow \O$.
\vskip.2cm

What about a generalization of the concept of a (pseudo-) Riemannian metric?
In the literature we find the following suggestions.
\begin{itemize}
\item Connes' distance formula \cite{Conn94} generalizes the Riemannian geodesic distance. This is an interesting new tool even in ordinary Riemannian geometry where, however, it is bound to the case of positive definite metrics (see \cite{Parf+Zapa98} for an attempt to overcome this restriction). Its generalization to noncommutative geometry requires some more understanding what a convenient counterpart to the Riemannian Dirac operator should be. This makes model building quite complicated.
\item In several papers a metric has been considered as an element of $\O^1 \oa \O^1$ or as a map $\O^1 \oa \O^1 \rightarrow \A$. In particular, a technical problem arose, namely the impossibility to extend a linear connection (on $\O^1$) to a connection on $\O^1 \oa \O^1$ in certain examples (see \cite{Mour95,BDMHS96}). Such an extension is needed in order to define metric compatibility in a straight way. Furthermore, according to our knowledge there have been no applications so far to really prove the usefulness of this definition of a metric in noncommutative geometry.
\item A generalization of the Hodge $\star$-operator to noncommutative geometry appeared to be a useful structure in some applications based on noncommutative geometry of commutative algebras \cite{DMHS93,DMH96_id,DMH97_im}. It has been generalized to noncommutative algebras
in \cite{DMH98_Prag}.  
\end{itemize}

Departing from all of these definitions, for (noncommutative) differential calculi on {\em commutative} algebras we propose that a metric should be taken to be an element of $\O^1 \oL \O^1$ where $\oL$ is the left-linear tensor product which satisfies
\be
    (f \, \alpha) \oL (h \, \beta) = f \, h \, \alpha \oL \beta
\ee
for $f,h \in \A$ and $\alpha, \beta \in \O^1$ (see also \cite{DMH99_dRg}). The next section provides some examples through which we were led to this definition of a metric. It should be noticed that there is no direct way to extend this definition to the case of noncommutative algebras $\A$. In some sense, however, the Hodge operator mentioned above may be regarded as such an extension.
\vskip.2cm

In the following we make use of the fact that, given a differential calculus on a commutative algebra $\A$, there is a unique associative and commutative product $\bullet$ in the space of 1-forms such that $\alpha \bullet \d f = [\alpha , f ]$ and
\be
    (f \alpha h) \bullet (f' \beta h') = f f' \, (\alpha \bullet \beta) \, h h'
    \qquad \forall f,f',h,h' \in \A, \, \alpha, \beta \in \O^1
\ee
\cite{BDMH95}. We call it the {\em canonical product} in $\O^1$. It measures the deviation from the ordinary differential calculus where $\alpha \bullet \beta = 0$ for all $\alpha, \beta \in \O^1$.

\section{Where left-linear pseudo-Riemannian metrics show up}
 
\subsection{Metrics on finite sets}
Let $\A$ be the algebra of functions on a finite set $M$. It has been shown in \cite{DMH94_ddm}
that first order differential calculi on $\A$ are in one-to-one correspondence with digraphs the vertices of which are the elements of $M$. Given such a digraph, we associate with an arrow from $i \in M$ to $j \in M$ an algebraic object $e^{ij}$.
Let $\O^1$ be the linear space (over $\Cx$) generated by all these $e^{ij}$.
An $\A$-bimodule structure can then be introduced via
\be 
    f \, e^{ij} \, h = f(i) \, e^{ij} \, h(j)  \qquad
    \forall f,h \in \A  \; .            
\ee     
With 
\be
   \d f = \sum_{i,j} [ f(j) - f(i) ] \, e^{ij}
\ee
we obtain a first order differential calculus. It is natural to regard the set of outgoing arrows at some point $i \in M$ as the analogue of the cotangent space in ordinary continuum differential geometry.
\vskip.2cm

As a candidate for a metric let us consider $g \in \O^1 \oa \O^1$. Using the above formulas and the properties of the tensor product over $\A$, we obtain
\be
    g = \sum_{i,j,k} g_{ijk} \, e^{ij} \oa e^{jk}
\ee    
with constants $g_{ijk}$. Here $e^{ij}$ and $e^{jk}$ live in different cotangent spaces
and it would be quite unnatural for a metric to compare vectors located at different points. In contrast, if we take $g \in \O^1 \oL \O^1$, then
\be
     g = \sum_{i,j,k} g(i)_{jk} \, e^{ij} \oL e^{ik} 
\ee
with constants $g(i)_{jk}$. Here $e^{ij}$ and $e^{ik}$ live in the same cotangent space and this enables us to make contact with classical geometry \cite{DMH99_dRg}.

\subsection{A class of noncommutative differential calculi on $\Rl^n$}
In terms of coordinates $x^\mu$, $\mu =1,\ldots,n$, on $\Rl^n$ a class of first order\footnote{A first order differential calculus $\d \, : \, \A \rightarrow \O^1$ 
always extends to higher orders via the rules of differential calculus.}  differential calculi is determined by the commutation relations\footnote{On the rhs we use the summation convention (summation over $\kappa$).}
\be
     [ \d x^\mu, x^\nu ] = \ell \, C^{\mu \nu}{}_\kappa \, \d x^\kappa
        \label{dc-C}
\ee
where $\ell$ is a constant and $C^{\mu \nu}{}_\kappa$ are functions of the coordinates which have to satisfy certain consistency conditions \cite{DMHS93,BDMH95}. In terms of the canonical product in $\O^1$ this becomes
$ \d x^\mu \bullet \d x^\nu = \ell \, C^{\mu \nu}{}_\kappa \, \d x^\kappa $.
We assume that $\d x^\mu$ forms a basis of $\O^1$ as a left- and as a right $\A$-module. Generalized partial (left- and right-) derivatives can then be introduced via
\be
   \d f = (\pa_{+\mu} f) \, \d x^\mu = \d x^\mu \, (\pa_{-\mu} f)  \; .
\ee
A coordinate transformation is a bijection $x^{\mu'}(x^\nu)$ such that 
$\pa_{+\nu} x^{\mu'}$ is invertible. Then
\bea 
     [ \d x^{\mu'}, x^{\nu'} ] 
 &=& \pa_{+\kappa} x^{\mu'} \, [ \d x^{\kappa}, x^{\nu'} ]
  =  \pa_{+\kappa} x^{\mu'} \, [ \d x^{\nu'}, x^\kappa ]     \nonumber \\
 &=& \pa_{+\kappa} x^{\mu'} \, \pa_{+\lambda} x^{\nu'} \, [ \d x^\lambda, x^\kappa ]
  = \ell \, \pa_{+\kappa} x^{\mu'} \, \pa_{+\lambda} x^{\nu'} \, 
    C^{\kappa \lambda}{}_\sigma \, \d x^\sigma
\eea 
using the commutativity of $\A$ and the derivation property of $\d$. Hence,
\be
    C^{\mu' \nu'}{}_{\kappa'} 
 = \ell \, \pa_{+\kappa} x^{\mu'} \, \pa_{+\lambda} x^{\nu'} \, 
    C^{\kappa \lambda}{}_\sigma \, \pa_{+\kappa'} x^{\sigma}  \;.
\ee
If we define
\be
   g^{\mu \nu} = C^{\mu \kappa}{}_\lambda \, C^{\lambda \nu}{}_\kappa
\ee
we obtain the transformation rule
\be
   g^{\mu' \nu'} = \pa_{+\kappa} x^{\mu'} \, \pa_{+\lambda} x^{\nu'} \,
                   g^{\kappa \lambda}  \; .
\ee
Suppose an inverse $g_{\mu \nu}$ exists. Then
\be
   g_{\mu' \nu'} = \pa_{+\mu'} x^\kappa \, \pa_{+\nu'} x^\lambda \,
                   g_{\kappa \lambda}  \; .
\ee
This is {\em not} compatible with $g = g_{\mu \nu} \, \d x^\mu \oa \d x^\nu$, but rather with 
\be
    g = g_{\mu \nu} \, \d x^\mu \oL \d x^\nu   \; .
\ee
We mention that one can prove that if $g^{\mu \nu}$ is invertible, then there is an $\A$-module basis $\theta^\mu$ of $\O^1$ such that
\be
   \theta^\mu \bullet \theta^\nu = \delta^\mu_\kappa \, \delta^\nu_\kappa \,
   \theta^\kappa  \; .
\ee
If the $\theta^\mu$ are holonomic, then we have $C^{\mu \nu}{}_\kappa = \delta^\mu_\kappa \, \delta^\nu_\kappa$ 
and therefore the lattice differential calculus considered in \cite{DMHS93}.

(\ref{dc-C}) is the basis of an interesting physical model. Think of the $x^\mu$ as space-time coordinates. If the functions $C^{\mu \nu}{}_\kappa$ and the $g^{\mu\nu}$ derived from them are dimensionless, then $\ell$ should have the dimension of a length and a natural candidate for it would be the Planck length. The kinematical structure of space-time is then modified at the Planck scale.

\subsection{From the Hodge operator to the metric tensor}
Let $\A$ be a {\em commutative} algebra and $(\O(\A), \d )$ a differential calculus over $\A$ which admits linear and invertible maps
\be
    \star \; : \; \O^r \rightarrow \O^{n-r}  \qquad  r=0,\ldots,n
\ee
for some $n \in \Nl$ such that\footnote{In order to generalize this to a {\em noncommutative}  algebra $\A$, an involution ${}^\ast$ on $\A$ is needed and the rhs has to be replaced by $f^\ast \star w$ \cite{DMH98_Prag}.}
\be
    \star \, (w \, f) = f \, \star \, w  \qquad  \forall f \in \A, \; w \in \O \; .
              \label{star-cov}
\ee
As a consequence, $\star^{-1} (f w) = (\star^{-1} w) \, f$. The set of maps $\star$ is called a (generalized) {\em Hodge operator}. It induces an inner product in $\O^1$ as follows,
\be
    (\alpha , \beta) = \star^{-1} (\alpha \star \beta)  \; .
\ee
As a consequence of (\ref{star-cov}), it satisfies
\be
     (\alpha , \beta \, f) = (\alpha \, f, \beta) \, , \quad
     (f \, \alpha , \beta) = f \, (\alpha , \beta) \;.
\ee
In applications of the formalism in the context of completely integrable models (and in particular generalized principal chiral models) \cite{DMH96_id,DMH97_im,DMH98_Prag}, a {\em symmetric} Hodge operator was needed, i.e.,
\be
    \alpha \star \beta = \beta \star \alpha \qquad  \forall \alpha, \beta \in \O^1 \;.
\ee 
As a consequence of this rather restrictive condition, we have
\be
    (\alpha , f \, \beta) = (f \, \beta, \alpha) = f \, (\alpha, \beta) \;.
\ee
 For a differential calculus of the kind considered in the previous subsection, one can introduce metric components 
\be
     g^{\mu\nu} = (\d x^\mu , \d x^\nu)  \; .
\ee
Using the above formulas, the effect of a coordinate transformation is
\be
   g^{\mu'\nu'} = \pa_\kappa  x^{\mu'} \, \pa_\lambda x^{\nu'} \, g^{\kappa \lambda} \; .
\ee
As a consequence, if $g^{\mu\nu}$ has an inverse $g_{\mu\nu}$, then 
\be
    g = g_{\mu\nu} \, \d x^\mu \oL \d x^\nu
\ee
is a tensor (but not $g = g_{\mu\nu} \, \d x^\mu \oa \d x^\nu$).

\section{The metric compatibility condition}
Let $(\O(\A), \d)$ be a differential calculus over a commutative algebra $\A$ and 
$\nabla$ a linear connection. Let us introduce the twist map
\be
    \tau (\alpha \oL \beta) = \beta \oL \alpha
\ee
and the map
\be
  \bullet ((\alpha \oa \gamma) \oL (\beta \oa \delta))
 = (\alpha \bullet \beta) \oa (\gamma \oL \delta)
\ee
where the canonical product in $\O^1$ enters on the rhs. Now we define\footnote{An expression like $\alpha \oa \beta \oL \gamma$ has to be read as $\alpha \oa (\beta \oL \gamma)$.}
\be
    \nabla(\alpha \oL \beta) = \nabla \alpha \oL \beta + (\mbox{id} \oa \tau)
    (\nabla \beta \oL \alpha) - \bullet (\nabla \alpha \oL \nabla \beta) \;.
\ee
It is easy to verify that this defines a left $\A$-module connection on
$\O^1 \oL \O^1$. Now we can impose the condition
\be
     \nabla g = 0
\ee
on an element $g \in \O^1 \oL \O^1$. If $g$ is a candidate for a metric, this condition 
generalizes the familiar metric compatibility condition of ordinary differential geometry.

\section{Conclusions}
We have proposed a definition of a metric tensor as an element of $\O^1 \oL \O^1$ for (noncommutative) differential calculi on commutative algebras and presented examples in which such a structure appears naturally. Furthermore, a corresponding compatibility condition with a linear connection has been formulated. In the particular case of differential calculi on discrete sets, these structures have been explored in \cite{DMH99_dRg}.

\vskip.2cm
\noindent
 F. M.-H. is grateful to C. Burdik for the opportunity to present 
the material of this paper at the $8^{th}$ Colloquium on {\it Quantum Groups 
and Integrable Systems}.

\end{document}